# TRANSPORTATION FATIGUE TESTING OF THE pHB650 POWER COUPLER ANTENNA FOR THE PIP-II PROJECT AT FERMILAB *

J. Helsper,[†], S. Chandrasekaran, J. Holzbauer, N. Solyak
FNAL, Batavia IL, 60510, USA

## Abstract

The PIP-II Project will see international shipment of cryomodules from Europe to the United States, and as such, the shocks which can occur during shipment pose a risk to the internal components. Of particular concern is the coupler ceramic window and surrounding brazes, which will see stresses during an excitation event. Since the antenna design is new, and because of the setback failure would create, a cyclic stress test was devised for the antenna. This paper presents the experimental methods, setup, and results of the test.

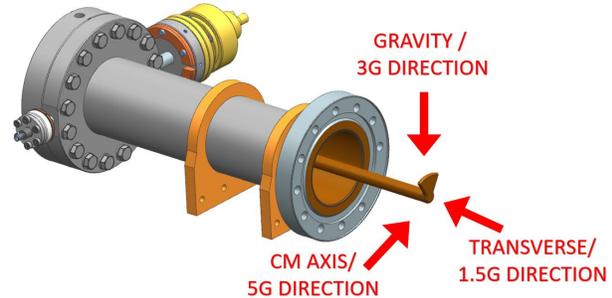

Figure 1: pHB650 Coupler Vacuum Side Assembly

## INTRODUCTION

The Proton Improvement Plan II (PIP II) is a multinational project which will overhaul Fermi National Accelerator Laboratory's (FNAL) linear accelerator (linac) [1]. FNAL will receive overseas shipments via land and air of cryomodules (CM), individual linac sections, from both the Commissariat à l'Énergie Atomique et aux Énergies (CEA) of France and the Science and Technology Facilities Council as part of the UK Research Innovations (STFC-UKRI) of the United Kingdom. These shipments pose significant risk to the CMs if not carefully considered.

The first CM to ship will be the prototype high-beta 650 MHz (HB650) CM [2], which was evaluated for the rigors of transportation analytically [3], and for which the shipment frame [4] has been tested [5]. The low-beta 650 MHz (LB650) CM [6] will also be transported overseas to FNAL.

The RF couplers [7] used on these CMs are of a novel design, having a single ceramic window brazed to thin copper sleeves isolating the beamline vacuum from atmosphere.

Given this, and that past CM shipments have seen failure occur at the couplers [8], a higher level of scrutiny has been applied to the design to ensure its readiness for transport as part of the CM shipments. The vacuum side coupler assembly is shown in Fig. 1.

## METHODOLOGY

To validate the 650 MHz coupler design (used on both HB650 and LB650 CMs), first the dynamic behavior was checked. Measurement of the antenna's first resonant frequency and estimation of mechanical damping, coupled with information on the transportation itself, allowed for an estimation of the total number of stress-inducing oscillations the coupler can experience. With this number, the antenna was then forcibly displaced to simulate the shocks experienced during transportation, in fully reversed cycles. The deformation was applied slowly to prevent any dynamic effects, and the stress created as a result was considered quasi-static (i.e. equivalent to that of a static-structural finite element analysis). The resonance which followed any excitation was assumed to take the form of a similar stress state, as the fundamental resonant mode is that of a simple cantilevered beam.

While not identical to the true excitation scenario (large, sudden shocks), displacement of the antenna creates a near identical stress state. The effects of multiple shocks occurring within the mechanical decay period of the antenna are excluded by this method, but accounted for in the high factor of safety (FOS) in the final number of completed cycles.

A shaker test, while valuable for certain components, was not selected for the coupler antenna since the excitation of the antenna can be difficult to control accurately and the amplitude of movement could quickly exceed any realistic values, leading to lasting and unnecessary damage.

## ANALYSIS

### Mechanical/Stress Response

The shipping frame [4] of the HB650 CM is designed to limit shocks to 3G vertical, 5G axial, and 1.5G transverse. The frame also isolates shocks above 10 Hz by at least 80%, relative to the input excitation. See Fig. 1 for the aforementioned directions. The 5G axial shocks, coupled with the effects of gravity (5G+1G) are considered to be the worst case scenario.

To determine the required applied displacement to mimic 5G+1G, an ANSYS® model of the antenna was used. The boundary conditions account for vacuum, and are shown along with the mesh in Fig. 2. Only solid hexahedral mesh was used in the high stress areas, and all objects had a minimum of three elements through the thickness.

---
* WORK SUPPORTED, IN PART, BY THE U.S. DEPARTMENT OF ENERGY, OFFICE OF SCIENCE, OFFICE OF HIGH ENERGY PHYSICS, UNDER U.S. DOE CONTRACT NO. DE-AC02-07CH11359.
† jhelsper@fnal.gov

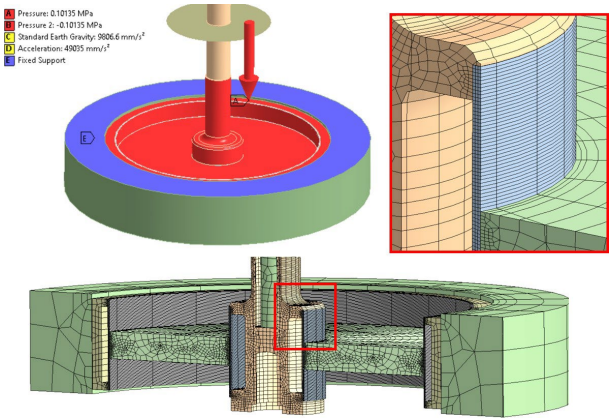

Figure 2: Analysis Boundary Conditions and Mesh

To mimic the antenna tester (discussed later on), displacement was applied approx. 21 mm below the antenna tip, as shown in Fig. 3. The applied displacement necessary to create the same stress state was 1.61 mm, which resulted in a displacement of 1.73 mm to be measured by the laser. The shape of the displacement (unit-less) is shown in Fig. 6, which is considered identical to the fundamental mode shape.

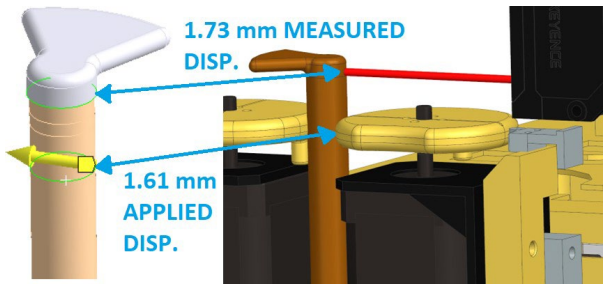

Figure 3: Applied and Measured Displacements

The mechanical stresses experienced by the coupler due to 5G + 1G for gravity is shown in Fig. 4. The same ANSYS® calculations were used to determine the amount of applied deformation needed to create a similar stress response, which was found to be approx. 1.73 mm.

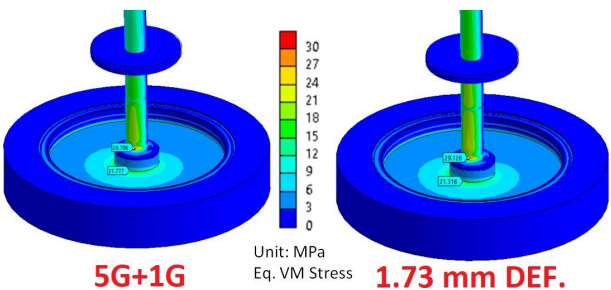

Figure 4: Stress Comparison A

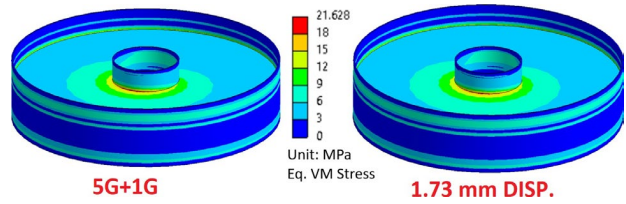

Figure 5: Stress Comparison B

## Dynamic Response

With the same boundary conditions and mesh shown in Fig. 2, the calculated first resonance of the antenna was 37 Hz, with the mode shape shown in Fig. 6.

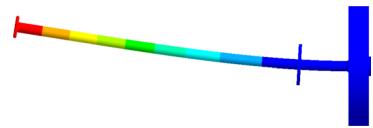

Figure 6: First Mode Shape / 5G+1G Displacement Shape

## RESONANCE VERIFICATION

### Methods

The first resonant frequency of the antenna was verified with two methods. Method A was to attach an accelerometer to the tip of the antenna, apply an excitation, and measure the response. Method B was to measure the displacement of the excited antenna with a high precision laser collecting data at >300 Hz. The setup used for both methods A and B is shown in Fig. 7.

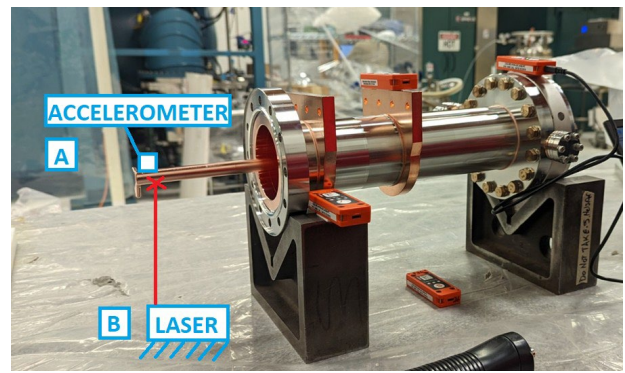

Figure 7: Resonance Verification A) By Accelerometer B) By Laser

### Results

As expected, the mass of the accelerometer lowered the fundamental frequency. With the additional mass, the resonance was predicted to be 29 Hz, and was measured as approx. 32 Hz, as shown in Fig. 8.

The resonance measured via laser was predicted to be 37 Hz, and was measured as 39 Hz, as shown in Fig. 9. The decay of excitation / mechanical damping effect can be seen in the plotted response of the antenna.

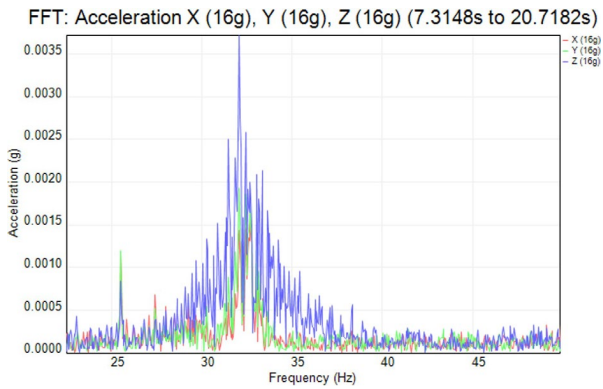

Figure 8: Accel. Distribution, Accelerometer

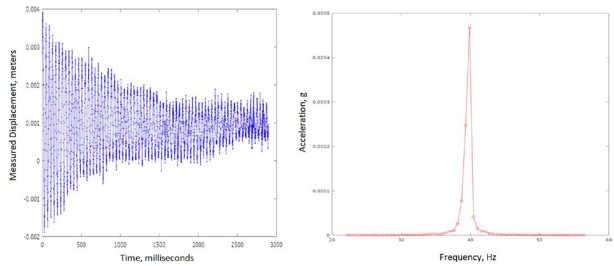

Figure 9: Laser Response and Accel. Distribution

Both results showed that the antenna was somewhat stiffer than predicted, but as the results were nominally close to the predictions, further investigation was not warranted.

## CYCLES ESTIMATION

To estimate the total possible shocks at 5G+1G equivalency the couplers could experience, the following information was necessary: total hours transported for each configuration; ring-down factor for excitation; distribution of shock magnitudes during shipment; effective damage of all shocks converted to 5G+1G equivalency.

### Hours Transported

Prior to assembly on the CM, the couplers will be shipped with the antenna in the vertical position, as vertical shocks are most prevalent, and this minimizes the stress to the coupler. The shipping duration from all possible vendors was considered, along with the need for warranty returns, delays during shipment, and shipment to partners for installation to the CM. A nominal shipment was estimated to be 120 hours total, while the need for a warranty return increased the value to 270 hours.

Post CM assembly, the antenna will be horizontal. The most any CM will be shipped overseas is twice, as is for the pHB650 CM, which was assembled at FNAL, and will be shipped round-trip to STFC-UKRI to validate the design. Including on-road testing before the trip, it was estimated the CM can see 120 hours of active transport.

### Ring Down Factor

Using the data shown in Fig. 9, it was estimated that it takes approx. 36 oscillations for the antenna to fully dampen its movement, or roughly 1 second considering the resonant frequency. As the oscillations decrease in amplitude, the stress induced decreases, and thus the damaging effect is also lessened. While in theory the ring down response should be exponential, it was considered as linear to be conservative in the cumulative damage estimates to the antenna.

### Shock Frequency and Distribution

The accelerator upgrade project (AUP), for which FNAL is collaborating with CERN, sees regular shipment via land and air over seas, and was representative of the potential shipments the couplers may see. The data set used was from shock logs on the outer crate (i.e. no damping). This is conservative as both individual coupler shipments and CM shipment will have some form of shock isolation. Actual data from CM shipment could not be used, as at the time of the tests, no HB650 shipments had occurred.

The count frequency for the vertical direction is shown in Fig 10. Due to the data sorting method used, shocks <1G were over represented, and corrected to be log-linear, which is conformal with the data trend and prior experience of the authors.

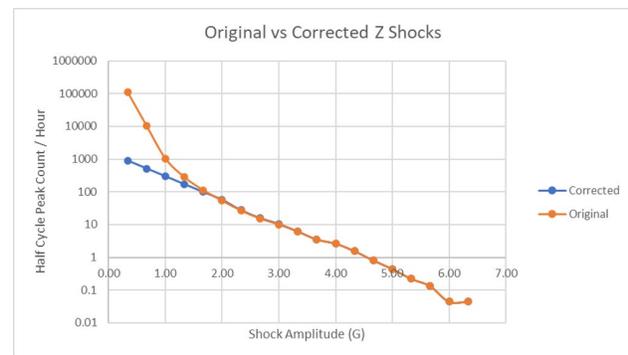

Figure 10: Vertical Shock Counts

### Effective Damage to Antenna

Based on orientations and associated hours of transport for the couplers, a table was generated tallying the counts for each shock magnitude for each principal direction. Shocks aligned with the coupler antenna axis were considered negligible.

Analytical fatigue calculations were used to determine the relative damage each magnitude and count of shock would have to the antenna. This was then converted into an equivalent number of 5G+1G displacements. Marin factors such as surface condition, size, load, temperature, etc. were based on the manufacturing, geometry, and operating conditions of the coupler. The desired failure rate was selected as <1/1000. The strength of copper was selected for the annealed state. This method found that the antenna should be tested to 42,200 fully reversed cycles at 1.73 mm displacement to have equivalent damage to what is predicted during all possible transportation. The table showing the summation of damage is shown below in Fig 11. This number of cycles was estimated to be 3% of the number of cycles required to experience failure.

| Shock Level, G | $\sigma_a$, max Stress, Mpa | Fully Rev. Cycles to Failure | Damage per cycle | Number of cycles predicted | Total damage |
|---|---|---|---|---|---|
| 0.33 | 7.6 | 5.65E+08 | 1.77E-09 | 3.58E+06 | 6.33E-03 |
| 0.67 | 9.5 | 2.28E+08 | 4.39E-09 | 1.38E+06 | 6.04E-03 |
| 1.00 | 11.4 | 1.08E+08 | 9.23E-09 | 5.53E+05 | 5.11E-03 |
| 1.33 | 13.2 | 5.78E+07 | 1.73E-08 | 2.36E+05 | 4.08E-03 |
| 1.67 | 15.1 | 3.35E+07 | 2.98E-08 | 1.07E+05 | 3.19E-03 |
| 2.00 | 17.0 | 2.07E+07 | 4.82E-08 | 4.99E+04 | 2.41E-03 |
| 2.33 | 18.9 | 1.35E+07 | 7.41E-08 | 1.98E+04 | 1.47E-03 |
| 2.67 | 20.8 | 9.14E+06 | 1.09E-07 | 1.14E+04 | 1.24E-03 |
| 3.00 | 22.7 | 6.41E+06 | 1.56E-07 | 7.51E+03 | 1.17E-03 |
| 3.33 | 24.6 | 4.62E+06 | 2.16E-07 | 3.95E+03 | 8.54E-04 |
| 3.67 | 26.5 | 3.42E+06 | 2.93E-07 | 2.11E+03 | 6.17E-04 |
| 4.00 | 28.3 | 2.58E+06 | 3.88E-07 | 1.57E+03 | 6.11E-04 |
| 4.33 | 30.2 | 1.98E+06 | 5.05E-07 | 9.33E+02 | 4.71E-04 |
| 4.67 | 32.1 | 1.55E+06 | 6.47E-07 | 4.80E+02 | 3.11E-04 |
| 5.00 | 34.0 | 1.22E+06 | 8.17E-07 | 2.67E+02 | 2.18E-04 |
| 5.33 | 36.5 | 9.14E+05 | 1.09E-06 | 1.33E+02 | 1.46E-04 |
| 5.67 | 39.0 | 6.96E+05 | 1.44E-06 | 8.00E+01 | 1.15E-04 |
| 6.00 | 41.5 | 5.39E+05 | 1.85E-06 | 2.67E+01 | 4.95E-05 |
| 6.33 | 44.1 | 4.24E+05 | 2.36E-06 | 2.67E+01 | 6.29E-05 |
| | 5G+1G Equivalent | | **4.22E+04** | Total Damage | 3% |

Figure 11: Cumulative Damage and Cycles Estimation

## FATIGUE TESTING

### Tester Design

The primary components of the fatigue tester are shown in Fig. 12. The tester was controlled by an Arduino Mega. A vacuum chamber served to mimic the transport stress conditions, and the vacuum gauge was monitored for change (indicating a leak). The chamber's upper and lower structure were made of aluminum structural channel. The mounting of the antenna is shown in Fig. 13. It was not possible to mount to the opposite side of the antenna as certain failure modes could not be detected.

The antenna was displaced by a 'snail' shaped cam lobe, which would rotate until the required displacement was achieved, and then return to its start position, verified by a limit switch. The cams were coated with PTFE to reduce wear, and had a safety stop to prevent excessive displacement in case of an unexpected event. Cams were actuated by stepper motors mounted to 1 Kg load cells, which recorded the force response of the antenna. Movement of the antenna was restricted to be uni-axial by a PTFE bar held snug to the antenna. Movement was monitored by a Keyence®IL-65 Laser. A non-reflective coating was applied to the antenna to prevent irregular measurements. The number of cycles completed, along with status indicators, were read out to an LCD display. The test could be started or stopped with a simple button.

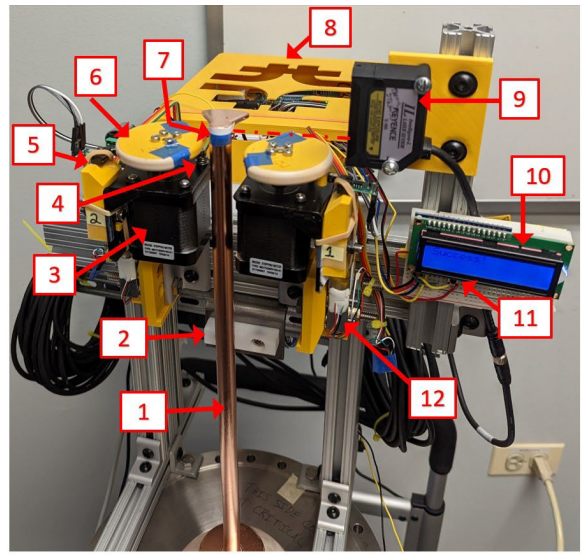

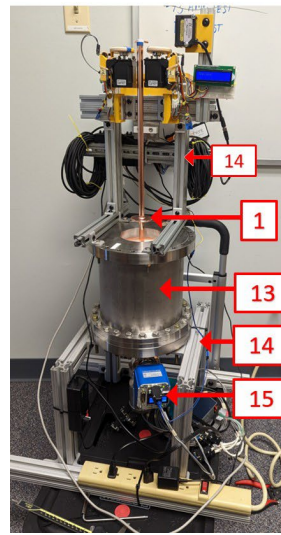

| Item | Description |
|---|---|
| 1 | pHB650 Antenna |
| 2 | PTFE Slider Support Bar |
| 3 | Stepper Motor |
| 4 | Cam Safety Stop |
| 5 | Roller Limit Switch |
| 6 | Cam Lobe, Snail Shape |
| 7 | Non-Reflective Covering |
| 8 | Controls Housing |
| 9 | High Precision Laser |
| 10 | LCD Readout |
| 11 | User Interface |
| 12 | Load Cell |
| 13 | Vacuum Chamber |
| 14 | Framing |
| 15 | Vacuum Gauge |

Figure 12: Antenna Tester Components

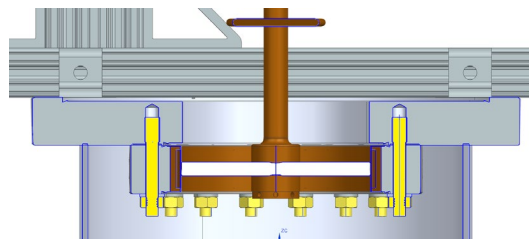

Figure 13: Mounting of Antenna, CAD Cross Section

### Tester Operation

One limitation of the Arduino Mega was the ability to only perform one task at a time, which is exemplified by the procedure below.

On startup, the tester would perform its initialization sequence. This confirmed the number of cycles to perform, and that the data collection laptop was connected. Then, each stepper motor would rotate incrementally until the desired displacement was reached. The load cell measurements were stored as reference values from this initial reading.

The tester would then perform 50 fully reversed displacement cycles, afterwards checking the displacement and load in each direction, along with vacuum levels. If the displacement varied by more than 2% from the ideal value, corrections would be performed, and testing would continue. If the measured load varied significantly from the initial value (>10%) the test would be halted. If the vacuum indicated a leak, the test would be halted. Once the tester reached the required number of cycles, the test was complete.

*Results*

In general, the antenna behaved as expected. The force response increased over the first 5,000 cycles due to the strain hardening of the annealed copper, and then plateaued. The vacuum level never changed, and the displacement was always maintained within 2% of the ideal 1.73 mm. Some variation in the measured load occurred, and is thought to be due to build up of PTFE particulate at the point of contact with the antenna. The force response in each direction for the duration of the test is shown in Fig. 14

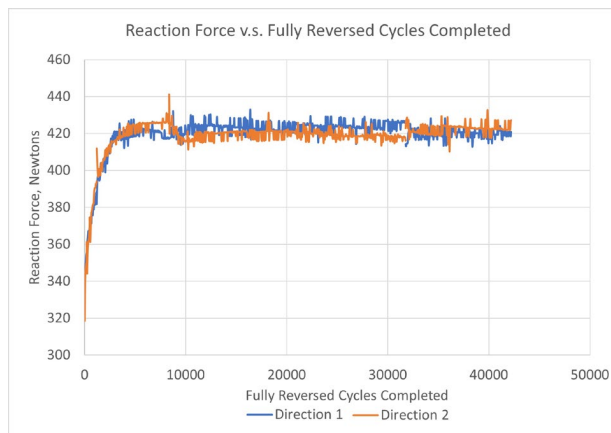

Figure 14: Results of Antenna Fatigue Test

## SUMMARY

In summary, the 650 MHz ceramic window antenna assembly has been evaluated and confirmed to be ready for the rigors of transportation, whether shipped by itself, or as part of the CM. This testing has given confidence that what has been commonly thought of as the 'weakest point' in the beamline vacuum barrier is capable of withstanding transportation.

## ACKNOWLEDGEMENTS

Special thanks to Maria Baldini for providing the AUP transport data used to estimate the number and magnitude of shocks the couplers will face during shipments.

## REFERENCES

[1] A.L. Klebaner, C. Boffo, S.K. Chandrasekaran, D. Passarelli, and G. Wu, "Proton Improvement Plan ' II: Overview of Progress in the Construction", in *Proc. SRF'21*, East Lansing, MI, USA, Jun.-Jul. 2021, pp. 182–189. doi:10.18429/JACoW-SRF2021-MOOFAV05

[2] V. Roger, N. Bazin, S.K. Chandrasekaran, S. Cheban, M. Chen, R. Cubizolles, *et al.*, "Design of the 650 MHz High Beta Prototype Cryomodule for PIP-II at Fermilab", in *Proc. SRF'21*, East Lansing, MI, USA, Jun.-Jul. 2021, pp. 671–676. doi:10.18429/JACoW-SRF2021-WEPTEV015

[3] J. Helsper, S. Cheban, "Transportation Analysis of the Fermilab High-Beta 650 MHz Cryomodule", in *Proc. SRF'21*, East Lansing, MI, USA, Jun.-Jul.2021, pp. 682-686. doi.10.18429/JACoW-SRF2021-WEPTEV017

[4] M. Kane, T. Jones, E. Jordan, J.P. Holzbauer, "Design of a Trasnport System for the PIP-II HB650 Cryomodule", in *Proc. LINAC'22*, Liverpool, UK, Jun.-Jul.2021, pp. 498-500. doi.10.18429/JACoW-LINAC2022-TUPOGE08

[5] J.P. Holzbauer, *et al.*, "Prototype HB650 Transportation validation for the PIP-II Project", in *Proc. LINAC'22*, Liverpool, UK, Jun.-Jul.2021, pp. 523-525. doi.10.18429/JACoW-LINAC2022-TUPOGE08

[6] N. Bazin, *et al.*, "Design of the PIP-II 650 MHz Low Beta Cryomodule", in *Proc. SRF'21*, East Lansing, MI, USA, Jun.-Jul.2021, pp. 841-844. doi.10.18429/JACoW-SRF2021-THPTEV006

[7] J. Helsper *et al.*, "Design, Manufacturing, Assembly, Testing, and Lessons Learned of the Prototype 650MHz Couplers", in *Proc. LINAC'22*, Liverpool, UK, July 2022, paper TUPOPA25, pp. 462-465. doi:10.18429/JACoW-LINAC2022-TUPOPA25

[8] N. Huque et al., "Improvements to LCLS-II Cryomodule Transportation" in *Proc. SRF '19*, Dresden, Germany, June 2019, paper TUP094, pp. 686-691.